# A Survey of fault mitigation techniques for multi-core architectures

Shashikiran Venkatesha and Ranjani Parthasarathi, Department of Information Science and Technology, College of Engineering Guindy Anna University, Chennai -600 025 Republic of India.
Email: shashikiran.annauniv@gmail.com and ranjani.parthasarathi@gmail.com

**Abstract**— Fault tolerance in multi-core architecture has attracted attention of research community for the past 20 years. Rapid improvements in the CMOS technology resulted in exponential growth of transistor density. It resulted in increased challenges for designing resilient multi-core architecture at the same pace. The article presents a survey of fault tolerant methods like fault detection, recovery, re-configurability and repair techniques for multi-core architectures. Salvaging at micro-architectural and architectural level are also discussed. Gamut of fault tolerant approaches discussed in this article have tangible improvements on the reliability of the multi-core architectures. Every concept in the seminal articles is examined with respect to relevant metrics like performance cost, area overhead, fault coverage, level of protection, detection latency and Mean Time To Failure. The existing literature is critically examined. New research directions in the form of new fault tolerant design alternatives for both homogeneous and heterogeneous multi-core architectures are presented. Brief on an analytical approach for fault tolerating model is suggested for Intel and AMD based modern homogeneous multi-core architecture are presented to enhance the understanding of the readers about the architecture with respect to performance degradation, memory access time and execution time.

## 1 INTRODUCTION

Fault tolerance: occurrence of faults in digital circuits could be due to bad electrical connections, broken components, burnt-out interconnect, chip logic error, radiation, electromagnetic interference and ground loops. Fault is an indicator of a "defect" at the device function level. Error is a visible outcome or result produced by a malfunctioning circuit or system. Further, faults can be classified as permanent faults and non-permanent faults. The permanent faults are caused by defects in the in the silicon or metals used for the processor package. The permanent faults are permanent and cause processor failure. Non-permanent faults can be classified into transient faults and intermittent faults. Cosmic rays, alpha particles, air pollution, humidity, electromagnetic interference and ground loops cause transient faults or soft errors. Manufacturing leftovers, incomplete oxide degradation, process variations and developing wear out are the major source for the intermittent faults appear periodically or aperiodically. Fault tolerance is interpreted as the capability of a system to provide normal service in the presence of faults. Error is a manifestation of fault in the system. Error detection and recovery are integral part of fault tolerance. Classical n-modular redundancy approaches have been used for detecting the errors at logical (combinational and sequential logic), micro-architectural and processor level. Triple Modular Redundancy (TMR) and Dual Modular Redundancy (DMR) methods are like special case of classical n-modular redundancy systems. Error recovery can be done in two ways, they are Forward Error Recovery (FER) (2) Backward Error Recovery (BER). FER techniques do not implement rollback to previous valid state to detect and correct the errors. In BER techniques, intermittently execute checkpoints to save present state, so that fault detection leads to automatic transition to previous valid state. Both FER and BER are well studied approaches.

Single core to multi-core: According to Moore's law, the number of transistors on a chip would double every 18 months. The performance scaling in single core has followed the trajectory laid down by Moore's law [1, 2]. 35% and 10% to 20% rise in the transistor density and die size respectively is reigned by Pollack's Rule [3], which states that performance increases as the square root of increase in complexity. Dennard law [4] stated that reducing the size of a transistor would improve circuit delay and consume less power. It is corroborated by the fact that Pentium 4 operated at 1.5GHz as compared to Intel i386 operated at 16MHz 20years back. Beyond 65 nm technology node, Dennard law failed, and it ignored leakage current and threshold voltage as it formed the baseline power for a transistor. As the size of the transistor reduced, it increased the power density and created a "power wall" that limited the processor frequency at 4 GHz in 2006. Instruction level parallelism (ILP) techniques like dynamic scheduling (scoreboard), with renaming (Tomasulo's approach), dynamic branch prediction, speculative execution and multiple instruction issuing improved the performance of the single core thereby reducing clock cycle per instruction. Simultaneous Multithreading (SMT) [5] introduced by Tullsen *et al.*, performs runtime resource sharing (e.g., register file, instruction queue, etc.) amongst hardware threads and are executed in parallel. The Intel Pentium 4's version of SMT, Hyperthreading [6], was first commercially available SMT processors with permissible two hardware threads at one time. Despite the gains, performance per watt /per mm die area was diminishing. Architects with diminishing prerogatives adapted divide and conquer approach resulting multi-core chip fabrication.

Fault tolerance in multi-core: In 1971, the Intel 4004 processor had 2,300 transistors and in early 2015 Intel released Xeon E5 2600 with 18 cores containing 5.56 billion transistors. Rapid advances in CMOS technology have resulted in an exponential increase in transistor density. This has led to many challenges such as increased power dissipation, thermal dissipation, occurrence of faults in the circuits and reliability issues. An 8 percent escalation in soft-error rate per logic state bit for each technology generation is anticipated by researchers [7]. The failure rate will be 100 times more at 16nm than at 180nm. Next to power dissipation, permanent and transient faults are emerging as the next major concern for architects designing reliable computer systems. Finally, fault tolerance has emerged as design characteristic like performance, power consumption, and temperature dissipation for reliable multi-core systems.

Considering above imperatives, the article presents a survey of error detection techniques at micro-architectural, core and thread level. Software based redundancy approaches, built in self-test methods, fault symptom detection approaches, reconfiguration and repair approaches, salvaging at micro-architectural and architectural level and error recovery techniques appeared in representative articles are discussed. Dimitris Gizopoulos *et al.* [8] has summarized error detection and recovery methods. The article [8] discusses micro-architectural redundancy methods for fault detection in lighter vein. Energy efficient redundant execution methods and time redundancy approaches are reviewed in our article. Many recent contributions are included in our article. We propose a classification based on the existing literature with new fault tolerant design alternatives (includes existing) for homogeneous and heterogeneous multicore system. Classification provides a preview of multicore architectures for researchers to evolve low-cost solutions with emphasis on power, Fault Tolerance (FT), Instruction Set Architecture (ISA) and reliability. We also present critical examination of existing literature and propose a quantitative approach for fault tolerating model proposed for modern homogeneous multicore architectures. The brief on quantitative analysis with respect to modern homogeneous multi-core architectures is to provide insight on performance loss, cache miss, memory access time and execution time (with respect to the architecture considered) and would also be helpful to simulation developers. The remaining portion of this article is organized as: review of the existing literature in section 2; Critical examination of existing literature in section 3; classification of multi-core architectures in section 4; analytical approach to fault tolerating model for modern homogeneous multi-core architectures in section 5.

## 2 REVIEW OF FAULT MITIGATION TECHNIQUES

The section presents a review on fault mitigation techniques that have tangible implications on the reliability of the multi-core system. Multi-core processor architecture comprises CPU cores, cache memory, register file, interconnection logic, and main memory. In the processor die, a major portion of the area is occupied by memory, which is protected by Error Correcting Codes (ECC). Fault tolerant designs are provided to protect the remaining portion of the die covering CPU and memory hierarchy control logic. Firstly, the fundamental methods (or approaches) adapted in the seminal articles are discussed in the section 2.1. Secondly, elaborate discussion on fault mitigations techniques advocated in seminal articles is presented in the sections 2.2 –2.6. Finally, critical examination of the existing techniques is presented in the section 3.

### 2.1 Fundamental methods of fault mitigation

The redundancy (spatial or temporal) method forms the basis for fault tolerance solutions for both fault detection and fault correction of permanent faults and transient faults. Table 1 details the approaches for detection and correction of transient and permanent faults with or without redundancy. The redundant components (like cores, threads, and instructions or software) and replicated functional units at pipeline level form the integral part for any fault detection or correction techniques. Error detection/correction and recovery together form fault tolerant solutions for any processor (uni/multi-core) architecture. Existing fault tolerance solutions can be conceptually classified as follows:

(a) Replicated structures/execution units as fault detectors/correctors,
(b) Cores as fault detectors/correctors,
(c) Threads as fault detectors/correctors,
(d) Replicated instructions/ software as fault detectors/correctors and
(e) Error recovery techniques.

**Table 1 Approaches for fault detection and correction.**

| Faults | Approach for Detection | Approach for Correction |
|---|---|---|
| Permanent Fault | Spatial redundancy or using invariants. | One or more redundant components that are assured to be correct. |
| Transient Fault | Any method that validates the correctness of program | Temporal or spatial re-execution of affected block of instructions. |

Each classification is briefly examined below.
(a) Replicated structures/execution units as fault detectors/ correctors

The replication of execution units and re-execution of instructions are the two principal approaches at the microarchitecture level. The replication of execution units in the pipeline was advocated by Ray *et al.* (2001) [9] & Parashar *et al.* (2004) [10]. The instruction re-execution strategy is incorporated in the proposals given by Nickel &

Somani (2001) [11] and Gomaa *et al.* (2005) [12]. Additional execution units such as checkers and low-cost adders are used in the approaches given by Shyam *et al.* (2006) [13] & Meixner *et al* (2007) [14] respectively. Hu *et al.* (2005) [15], provided an approach that exploited inherent structural redundancy by using idle time of floating-point unit for verifying the results of integer unit. Soman *et al.* (2015) [16] proposed an alternative to traditional approaches by incorporating additional hardware unit named Instruction Re-executions Units (IRUs) that is shared among '*n*' cores on the chip.

(b)  Cores as fault detectors/correctors.

Fundamental approach for detecting and correcting permanent faults at Core level would be N-modular redundancy technique. Dual Modular Redundancy (DMR) approach and Triple Modular Redundancy (TMR) are two major approaches commonly used for fault detection and fault correction. DMR and TMR approaches are realized at core level as tight lock-step and loose lock-step redundant execution approaches. In the tight lock-step redundant execution approach, cores execute the same instruction and compare the outputs cycle by cycle to detect errors. In loose lock-step redundant execution approach, cores are not synchronized and execute the same instruction. Recent contributions propose small light weight cores contextually named as checkers, verifiers or trailing cores, used to check the results of main core for fault detection. The proposals that adopt loose lock-step redundant execution are given by Bernick *et al.* (2005) [17], Austin *et al.* (1999) [18], Purser *et al.* (2000) [19] and Rashid *et al.* (2005) [20]. Li *et al.* (2017) [21] proposed N-Modular Redundancy (NMR) approach tacitly implemented by taking advantage of inherent core redundancy that exist in multi/many cores system. Iturbe *et al.* (2017) [22] proposed the use of loose lock-step TMR approach in Triple Core Lock Step (TCLS) processor in ARM Cortex R5 CPU to address the soft error vulnerabilities. Ainsworth and Jones (2018) [23] suggest a core level redundancy approach in multi-core ecosystem consisting out-of-order cores with groups of low power in-order small cores used for fault detection.

(c)  Threads as fault detectors/correctors.

The multiple threads executing on a physical core do not have much differences in their architectural state and hence inter-thread communication is faster than inter-core communication. IBM G5 (Spainhower & Gregg 1999) [24] is the first commercial machine that have thread level redundancy with two separate pipelines running simultaneously with an extra pipeline stage to check the results. The Redundant multithreading (RMT) methods advocated by Rotenberg *et al.* (1999) [25], Reinhardt *et al.* (2000) [26], Vijaykumar *et al.* (2002) [27] and Gomaa *et al.* (2003) [28] use two threads executing the same instructions on chip multiprocessors (CMPs). Huang *et al.* (2014) [29] suggested a hybrid model that incorporates RMT on multiple cores each having different performance levels.

(d)  Replicated instructions/software as fault detectors / correctors

Symptom based fault detection, instruction replication, monitoring dataflow and control flow invariants are the techniques adopted at the software level. The idea to replicate instruction execution is given by Oh *et al.* (2002a) [30]. Software based approaches for monitoring data and control flow invariants are proposed by Kanawati *et al.* (1996) [31], Reis *et al.* (2005) [32] and Meixner *et al.* (2008) [33]. Detecting and correcting transient faults by monitoring anomalies like page fault, hardware traps etc. is proposed by Li *et al.* (2008b) [34]. Hari *et al.* (2009) [35] extended the idea of symptom-based fault detection to multi-core systems. Nitin *et al.* (2015) [36] proposed a value locality-based scheme that uses state machine to monitor bit flips in value. Liu *et al.* (2017) [37] proposed selective instruction duplication (DMR at instruction level) and it relies on sensor based soft error detection approach given by Upasani *et al.* (2014) [38]. A novel software testing-based approach given by Mahmoud *et al.* (2019) [39] has a higher speed up in resilience analysis accompanied by accuracy in error detection.

(e)  Error recovery techniques

The data errors in the multi-core scenario occur due to following reasons:

(e.1) inappropriate synchronization between processors,

(e.2) dependencies between threads not handled properly and

(e.3) incorrect instruction sequence execution.

Error recovery techniques have significant role in ensuring correctness of the program execution in multi-core systems. Error recovery techniques can be classified into (1) Forward Error Recovery (FER) Techniques and (2) Backward Error Recovery (BER) Techniques. The FER continues to move forward in presence of errors and uses redundancy (TMR) to mask the failures. It also incorporates error detecting and correcting codes. FER does not rollback to previous fault free state. BER uses check points and logging mechanisms to recover from failures. BER rolls back to the latest validated checkpoint when a fault is detected. Seminal BER approaches for multi-core scenario are given by Sorin *et al.* (2002) [40], Prvulovic *et al.* (2002) [41], Nakano *et al.* (2006) [42], Dondalis & Prvulovic (2012) [43] and Rish *et al.* (2011) [44]. Sustained error recovery as the number of cores increase would be a significant requirement and one such solution is given by Rish *et al.* (2011) [44].

Considering the above conceptual classification, discussion in the seminal articles can be broadly classified as:

2.2 Fault detection techniques,

2.3 Reconfiguration and repair techniques,

2.4 Micro-architectural core salvaging techniques,

2.5 Architectural core salvaging techniques and

2.6 Error recovery and repair techniques.

Each of these techniques is discussed below.

**2.2 Fault Detection Techniques**

The existing literature on fault detection methods can be broadly divided into three main sections as:

2.2.1 Fault detection by redundancy

2.2.2 Built-In Self-Test (BIST) methods and

2.2.3 Symptoms-based fault detection methods.

**2.2.1 Fault Detection by Redundancy**

Fault detection by redundant operations can be classified into four classes of techniques and they are:

2.2.1.1 Core level techniques,

2.2.1.2 Micro-architectural level techniques,

2.2.1.3 Thread level techniques and

2.2.1.4 Software based redundancy techniques.

Each of these techniques is discussed below.

**2.2.1.1 Core level techniques**

The core level techniques are broadly classified as

- N-modular redundancy methods,
- Tight lock-step execution methods,
- Loose lock-step methods or loosely coupled processors,
- Signature based approaches,
- Dynamic verification methods,
- Fault isolation methods and
- Critical value forwarding methods.

TMR and DMR methods are special cases of classical N-modular redundancy systems, used to detect and correct faults or just detect faults. In tight lock-step redundant execution, two cores execute the same instruction and compare the outputs cycle by cycle to detect errors. The architectural state of the two cores must be the same and is achievable by identical initialization. The signal propagation latency in cores executing lock-step mode is not constant (Sarangi *et al.* 2006) [45]. Lock-step machines are not suitable for commercial market. They are suitable for mission critical systems [46,47].

In loose lock-step redundant execution, cores execute the same instruction but they are not synchronized. Non-Stop Advanced Architecture (NSAA) [17] and configurable isolation [48], which are closely related approaches follow loose lock-step redundant execution to detect faults. The architecture of NSAA has three groups of 4-way Itanium server processors. A core from each group forms a slice, each executing the same program. But they use separate address space. In NSAA, faults are inhibited at socket level and prevented from further propagation. Pairs of processes are handled simultaneously by NSAA. The configurable isolation [48] proposed by Aggarwal *et al.*, has a provision for reconfiguration of shared micro-architectural components like interconnects, caches and memory controllers when stuck by hard errors.

Fingerprinting technique [49] introduced by Smolens *et al.*, is a signature-based approach which uses Cyclic Redundancy Code (CRC) to create hash for the state of the processor comprising register updates, branch predictors, and load/store addresses. Reunion [50] is also a fingerprinting technique proposed by Smolens *et al.*, which has two cores - a vocal core and a mute core. Same thread is executed on two cores of the Chip Multi-Processor (CMP) architecture. The fingerprints are compared by the core before the memory write is visible to other cores. Two cores that do not replicate load values strictly will not have same view of the memory, unlike in the tight lockstep cores and this approach is called as relaxed input replication. By adopting relaxed input replication, Reunion incurs performance penalty of 5% and 6% for commercial and scientific workloads respectively. In strict model of replication, Reunion incurs an average performance penalty of 5% and 2% for commercial and scientific workload respectively.

In the fault isolation methods, during the interval between the occurrence of fault and the time at which it is detected, propagation of faults is not allowed to be visible outside the framework of programs. Dynamic Core Coupling (DCC) [51] approach proposed by LaFrieda *et al.*, performs fault isolation. In DCC, the processors compare results with each other arbitrarily in a DMR setup and avoid permanent core binding. This approach detects and recovers from both permanent faults and soft errors. It adopts the TMR approach to implement recovery procedure. Results presented show that DCC performs better than static DMR schemes and has 5% performance overhead for data mining applications and scalable scientific benchmark programs tested with eight threads on 16 cores.

Dynamic Instruction Verification Architecture (DIVA) [18] proposed by Austin *et al.*, is a runtime approach, which uses checker cores to monitor invariants by performing dynamic verification. The primary core implements all the normal functions of a conventional processor except loads and stores. The primary core executes "out of order" and the checker core verifies the results from primary core "in order". The checker core has been added to the commit phase of the processor pipeline and has an area overhead of 6% in ALPHA21264 processor. Slipstream [52] is similar to DIVA, but the two cores have identical microarchitecture. In DIVA, two cores are not identical. Slipstream has 100% hardware overhead and has 12% speedup compared to DIVA.

In the Critical value forwarding method, all values from a leading core are not forwarded to the trailing core. Redundant Execution using Simple Execution Assistance (RESEA) [53,54], is a technique that uses trailing cores for a subset of operations. The leading core sends only load values and branch decisions to the trailing core, thereby reducing bandwidth on interconnects and saves energy in the trailing core. Compared with non-redundant approaches, RESEA has <1% performance overhead and 1.34 times more energy consumption. Every two cycles, RESEA forwards one value from the leading core to the trailing core. Redundant Execution using Critical Value Forwarding (RECVF) [55] proposed by Subramanyan *et al.*, is a technique that forwards the results of the instruction on critical path of execution from leading core to trailing core. RECVF incurs low bandwidth cost on the interconnects, because it forwards only one third of normal transfer (load/store values, branch predictors) to trailing core. Compared to non-redundant approaches, RECVF has 1% performance overhead and 1.26 times more energy consumption. It has become attractive for adaption in Network on Chip (NoC) fabrics due to lesser bandwidth usage on interconnects. Both RESEA and RECVF are energy efficient redundant execution approaches and consume less energy compared to other existing approaches.

REMORA [56] proposed by Gopalakrishnan & Singh (2017), is a hybrid approach. It is a software method used to generate and insert static instruction signatures (SIS) as the first instruction in every basic block in the program binary. The dynamic instruction signatures are computed in the instruction decoder for every byte it decodes. When instruction decoder encounters the SIS instruction, it compares with dynamic instruction signature and detects faults if the results do not match. This safeguards only in-order fetch and decoder unit. Additional hardware unit called verifier; a light weight processor re-executes the instructions when they reach the head of Re-Order Buffer (ROB). All instructions in the verifier are independent i.e., data dependencies are resolved and the verifier can execute the instructions out-of-order. REMORA provides 100% fault coverage for transient faults. REMORA has code overhead, hardware overhead, performance penalty and power overhead of 3%, 6%, 28% and

**Table 2 Summary of error detection techniques at core level**

| Concept and year | Performance cost | Detection latency | Protects | Source of failure | Detection coverage | Area overhead |
|---|---|---|---|---|---|---|
| DIVA (1999) | Low (5%) | Low | Back end of core | Transient and permanent faults | 100% | 6% |
| Slipstream (2000) | 12% speedup | Deterministic | core | Transient and permanent faults | Very high | High |
| Lockstepping [(1998,2004)] | 1.5 – 2x | Cycle by cycle detection latency | core | Transient and permanent faults | 100% | 100% hardware cost in DMR based system |
| Fingerprinting (2004) | Low | Very high | (State) core | Transient and permanent faults | 100% | <100% |
| Reunion (2006) | 5%-9% | Low | core | Transient and permanent faults | Better than RMT | 100% |
| DCC (2007) | 3%-5% | Cycle by cycle detection latency | core | Transient and permanent faults | 100% | 100% + age table |
| RESEA (2009)(2010) | 1% | Deterministic | core | Transient and permanent faults | Less than 100% | Low |
| RECVF (2010) | 1% | Deterministic | core | Transient and permanent faults | Less than 100% | Low |
| REMORA(2017) | 28% | Deterministic | core | Transient faults | 100% | Low |
| PreFix(2017) | Low | Deterministic | core | permanent faults | Less than 100% | 1-4% |
| PED-HC(2018) | Low | Deterministic More than lockstep methods | core | Transient faults and permanent faults | Less than 100% | Low |

13% respectively.

PreFix [57], proposed by Soman & Jones (2017), is a technique that predicts instructions that are likely to use the faulty components. Faulty components are not turned off in this method. PreFix model consists of two types of cores: faulty cores and remote cores. Healthy cores are named as remote cores and cores with one or more faults are called as faulty cores. Every instruction is classified into one of the categories: "not faulty (NF)", "highly likely to fault" (HLF, the default), or "low likelihood of fault" (LLF) by the PreFix Predictor. NF instructions are not duplicated. HLF instructions are sent to remote cores for re-execution. If fault detectors catch a fault from LLF instructions, they are re-executed in remote cores. The power and area overheads for PreFix framework on 2-core machine are 3.5% and 1-4% respectively. Parallel Error Detection Using Heterogeneous Cores (PED-HC) [58] proposes a multicore architecture consisting of conventional out-of-order cores and small low power checker cores. Both checker cores and main core implement the same Instruction Set Architecture (ISA) and share the same address space in the memory. Main instruction sequence is split into small instruction sequence and re-executed on checker cores. Small instruction sequences are not data-dependent, enabling parallel error detection in checker cores. This approach can detect both soft and hard errors. Error detection latency is more than lock-step machines. The average error detection delay is 770ns and all loads and stores are checked within 5000ns for all evaluated benchmarks. The concepts are itemized and summarized in the table (Table 2).

### 2.2.1.2 Micro-architectural level techniques

At the micro-architectural level, both structural duplication approaches and time redundancy methods are available to detect the faults. Conceptually, the superscalar data path is used for detecting faults either by replicating a stage in a pipeline or re-executing a portion of a program during idle cycles of the pipeline.

Ray *et al*. [9] & Parashar *et al*. [10] have proposed the replication of the execution units. The instruction is dispatched to execution units from the renaming unit and the results are compared by hardware structures. If the results do not match, occurrence of hard faults in execution units can be inferred. Partial Explicit Redundancy and Implicit Redundancy Through Reuse (PER-IRTR) [12], is a time redundancy approach proposed by Gomma & Vijaykumar (2005), which introduces re-execution of portions or subsets of the program during the idle cycles of the pipeline. Explicit redundancy is provided only in low Instruction Level Parallelism (ILP) phase. For example, during cache miss, the redundant instructions are executed to give protection from soft error. A reduction in Soft Error Rate (SER) of 56% is observed in PER-IRTR. ARGUS [14] proposed by Meixner *et al*. (2007) is a structural duplication approach that monitors control-flow, computation, data-flow and memory correctness at run time in simple cores. It has an area overhead of 17% and performance overhead of 3.2-3.9% with fault coverage of 98% for permanent and transient faults. BulletProof [13] proposed by Shyam *et al*. (2006), protects the pipeline and the on-chip memory. A checker unit is connected with every stage of the pipeline like instruction decoder, arithmetic logic unit (ALU), register file and cache. A 9-bit mini-ALU is used for checking the ALU. BulletProof has 89% coverage for silicon material defects with a small area cost of 5.8%.

SHared REsource Checker (SHREC) [59] proposed by Smolens *et al*. (2004), is a time redundancy approach that moves instructions in program order from the Re-Order Buffer (ROB) into a small in-order queue for re-execution. The SHREC approach performs asymmetric and staggered redundant execution like PER-IRTR. The SHREC microarchitecture reduces the thread contention on shared resources such as issue queue and ROB by performing asymmetric redundant execution. SHREC provides soft error protection for the backend of a core. SHREC reduces performance penalty by 4% and 15 % for integer and floating-point applications respectively, when compared to well-designed superscalar architectures. Selective Replication (SeIR) [60] proposed by Vera *et al.* (2009), relies on Architecture Vulnerability Factor (AVF) to replicate instructions in the issue queue. Architecture Vulnerability Factor [61] is a metric that captures the fraction of bit transitions in registers or logic that could lead to visible errors. SeIR selectively re-executes smaller number of instructions above vulnerability threshold to attain larger coverage, thereby reducing the pressure on ROB. Both SHREC and SeIR are suitable for detecting and correcting soft errors only.

**Table 3 Summary of error detection techniques at micro-architectural level**

| Concept and year | Performance overhead | Detection latency | Protects | Faults target | Detection Mechanism |
|---|---|---|---|---|---|
| SHREC (2004) | Medium | Bounded | Backend of core | Transient faults | Re-execution in idle cycles |
| PER – IRTR (2005) | 2% | Bounded | Core | Transient faults | Re -execution |
| BulletProof (2006) | 5-25% | Bounded | Core | Permanent faults | BIST |
| ARGUS (2007) | 4% | Low | Core | Permanent and Transient faults | Monitoring invariants |
| SeIR (2010) | Low | Bounded | Backend of core | Transient faults | AVF estimation |
| REPAIR (2015) | Low | Bounded | Frontend of Core | Permanent faults | Re-execution |

REPAIR [16], a technique proposed by Soman *et al.* (2015) introduces Instruction Re-execution unit (IRU), a hardware unit shared among 'n' cores on-chip. REPAIR focuses on hard error detection. IRU executes instructions and services only one core at a time. Detection of faults are performed by detection/correction unit (DCU) attached to each core. A Core interface (CI) in REPAIR schedules instructions from main cores to IRU and follows round robin policy when multiple requests arrive at IRU. Experiments on four cores with four faults, one in each core, indicate that performance of REPAIR reduces by 47%. The concepts are itemized and summarized in the table (Table 3).

#### 2.2.1.3 Thread level techniques

The design of multithreading techniques to detect faults is less complex when compared with core-based techniques. The parameters like clock skew and propagation delay in buses do not affect threads communicating intermediate results. Threading provides reliability with less hardware cost. In 1995, IBM G5 [24] was the first commercial machine to incorporate multithreading by running two pipelines simultaneously. To check the results of two pipelines, one additional hardware stage was added. The task of the extra hardware stage is to compare the results of the two pipelines and infer faults, if the outcomes are different.

All the approaches discussed in this section use two threads to detect the faults and are loosely synchronized. Active stream / Redundant stream Simultaneous Multithreading (AR-SMT) [25] proposed by Rotenberg *et al.* (1999), is a technique which uses two threads 'A' and 'R'. Both the threads execute the same application. The 'A' thread runs tens of cycles in advance of the 'R' thread. The instructions are committed only if the results match. Threads perform reads and writes to different address spaces in memory. This approach

| Table 4 Summary of error detection techniques at thread level | | | | | | |
|---|---|---|---|---|---|---|
| **Concept and year** | **Performance cost** | **Detection latency** | **Protects** | **Sources of failure** | **Detection coverage** | **Area overhead** |
| AR – SMT (1999) | Very high | Hundreds of cycles | core | Transient faults | <100% | >2x |
| SRT (2000) | High | Unbounded | core | Transient faults | <100% | >2x |
| SRTR (2002) | Very high | Hundreds of cycles | core | Transient faults | <100% | x |
| CRT (2002) | CRT: achieves 13% better performance than a dual lockstep CPU | Unbounded | core | Transient and permanent faults | <100% | >2x |
| CRTR (2003) | Very high | 30 cycles | core | Transient and permanent faults | <100% | Very high |
| SlicK (2006) | Medium | Unbounded | core | Transient faults | <100% | Very high |
| SpecIV (2008) | Medium | Unbounded | core | Transient faults | <100% | Very high |
| PyDac (2014) | Medium | Unbounded | core | Transient faults | <100% | Very high |

incorporates value prediction and branch prediction techniques as adopted in Slipstream to accelerate the execution of both threads. This approach detects transient faults and corrects them with 10-30% overhead.

Simultaneously and Redundantly Threaded (SRT) processors [26], conceptualized by Reinhardt & Mukherjee (2000), is a technique that extends the idea of AR-SMT, by dynamically scheduling the instructions from redundant threads, which improves performance by 16% when compared to lock-step approaches. Vijaykumar *et al*. (2002) proposed SRT with soft error recovery capabilities which is called Simultaneously and Redundantly Threaded processors with Recovery (SRTR) [27]. The SRTR approach allows the results of the instructions executed by the leading thread to commit before verification which results in the system state being modified. Only outputs in the registers are checked and verified thereby increasing pressure on registers. SRTR approach can only detect and recover from soft errors. SRT fairs better than SRTR by 26% in terms of performance.

Redundant Multithreading (RMT) for single Simultaneous MultiThreaded (SMT) processor conceptualized by Mukherjee *et al*. (2002) and implemented on Chip MultiProcessors (CMP) is termed as Chip level Redundant Threading (CRT) [62]. This approach provides 13% better performance than lock-step cores with detection latency of 10 cycles or more. The Chip-level Redundantly Threaded multiprocessor with Recovery (CRTR) [28] proposed by Gomaa *et al*. (2003) is a technique that augments CRT for transient fault detection. CRTR maintains long duration between the leading thread and trailing thread, and the commit executed by it is asymmetric. Results of the trailing thread are used for recovery. CRTR incurs a performance overhead varying between 10-60% as compared to CMP (baseline architecture).

Slice-Based Locality Exploitation for Efficient Redundant Multithreading shortly referred as Slice Kill (SlicK) [63] proposed by Parashar *et al*. (2006), is a technique that relies on branch predictors, and the leading thread is executed in a full-fledged manner. SlicK leverages value and control flow locality to avoid full redundancy, thereby adopting partial redundancy approach on slices whose outputs are predictable. Instructions that belong to the backward slices or outputs of the instructions that are not predictable are re-executed by the trailing thread. Speculative Instruction Validation (SpecIV) [64] proposed by Kumar & Aggarwal (2008), is a technique that extends the basic principle of value prediction to all instructions in the leading thread. SpecIV incurs large overhead in area and is not a highly reliable solution.

PyDac [65] proposed by Huang *et al*. (2014), is a Python based task parallel programming model, which generates as many numbers of parallel tasks for the given application. The programming model utilizes divide and conquer concept from functional languages for task creation and data decomposition. The runtime system of this model reports errors from unreliable cores and re-executes those using redundant threads (DMR, TMR). The programming model and runtime system are evaluated on a 'green-white' architecture, where white cores are optimized for single thread performance and green cores are simple cores for high performance. This model uses dual threads to detect soft errors and triple threads to correct them. The runtime overheads in this model are largely dependent on the computation time of green and white cores and vary between 100-200% for evaluated benchmarks. The concepts are itemized and summarized in the table (Table 4).

### 2.2.1.4 Software based redundancy techniques

In order to decrease the hardware cost, time redundancy approaches have gained momentum. Conceptually, computations are repeated to detect faults by comparing the results. The software-based redundancy techniques can be broadly classified as

- Control flow checkers technique,
- Instruction duplication technique,
- Compiler based technique and
- Code triplication technique.

In the control flow checkers technique, only control statements are evaluated to detect errors. In the instruction duplication approach, every instruction is re-executed to detect faults. In the compiler-based approach, threads are created and assigned to cores by the compiler. The outcomes of the threads are compared to detect faults. In code triplication, two copies of the same program are executed and the third one uses AN code. In 'AN' code, A is a constant chosen arbitrarily and N is the word. The results from executing two copies of the same program are compared with an outcome from the computation using AN code, to detect and correct soft errors.

Signatured Instruction Streams (SIS) [66] proposed by Schuette & Shen (1987), Path Signature Analysis (PSA) [67] proposed by Namjoo *et al*. (1982) and Continuous Signature Monitoring (CSM) [68] proposed by Wilken and Shen (1990) are hybrid control flow checkers to detect the errors in the control flow at the fetch and decode stages. These methods have very high detection latency and instructions can change the architecture state before they graduate. Control Flow Checking by Software Signatures (CFCSS) [69] proposed by Oh *et al*. (2002b) is a software-based control flow checker. CFCSS ensures proper transfer of control to correct successor basic blocks. The exact direction of conditional branch instruction is not assured and this approach may not be suitable for superscalar processors.

Table 5 Software based error detection techniques

| Concept and year | Performance cost | Detection latency | Protects | Sources of failure | Detection coverage | Recovery |
|---|---|---|---|---|---|---|
| SIS, CSA, and CSM (1987,1982,1990) | Low | Unbounded | Control flow logic | Transient and permanent fault | Control flow errors | No |
| CFCSS (2002) | Low | Unbounded | Control flow logic | Transient and Permanent fault | Control flow errors | No |
| EDDI (2002) | >150% | Low and unbounded | core | Transient fault | 98.5% | No |
| SWIFT (2005) | Very high | Low and unbounded | core | Transient fault | <100% | No |
| CRAFT (2005) | Very high | Low and unbounded | core | Transient fault | <100% | No |
| SWIFT-R and TRUMP (2006) | Extremely high | Unbounded | core | Transient and permanent fault | <100% | Yes |
| SRMT (2007) | Very high | Unbounded | core | Transient fault | <100% | No |
| CLOVER (2017) | Varies with size of tail region | Low | core | Transient fault | <100% | yes |

Error Detection by Duplicated Instruction (EDDI) [70], SoftWare Implemented Fault Tolerance (SWIFT) [32], and CompileR Assisted Fault Tolerance (CRAFT) [71] perform instruction duplication. EDDI proposed by Oh *et al.* (2002a) is a software-based time redundancy approach. In EDDI, duplicate instructions inserted by the compiler, result in 100% performance overhead. EDDI and SWIFT proposed by Reis *et al.* (2005), are applicable for single threaded applications on both single and multicore processors. EDDI relies on compiler to create two redundant execution sequences for a single thread. Both the redundant execution sequences share register file and address spaces in the memory. EDDI incurs >111% performance overhead. EDDI and SWIFT are similar approaches. Store instruction is not protected by SWIFT, unlike EDDI. For example, the store instruction residing in the store buffer can be corrupted even when correct inputs are received. SWIFT assumes that memory is protected by ECC. SWIFT is an extension of CFCSS and assures that correct decisions of branch instructions are taken.

CRAFT [71] is a hybrid solution proposed by Reis *et al.* (2005) that addresses the inconsistency in SWIFT by integrating hardware structures to the existing solution. CFAFT protects load and store instructions or operations. The special buffer for load and store holds address and data relevant for write/read operation. The replica load and store instructions access this buffer for checking. The buffer finally commits the data to memory or register. Wang *et al.* (2007) introduced Software-based Redundant Multi-Threading (SRMT) [72], which is a compiler-based method for transient fault detection. The redundant threads are created by the compiler and can suitably be executed on the CMP systems. The threads communicate through reserved memory space. No hardware extension is needed unlike in SRT, SRTR, CRT and CRTR.

SWIFT-R [73] proposed by Chang *et al.* (2006), is a technique that provides recovery unlike all the above (SIS to SRMT). It works on the principle of majority voting by linking three copies of the same program. It protects load and store instructions. TRUMP (Triple Redundancy Multiplication Protection) provides error detection and correction using AN-codes. It executes two copies of the same program and one uses AN-codes. They detect both permanent and transient faults. SWIFT-R uses software RMT to provide forward error recovery. SWIFT-R incurs performance penalty of >200% due to triplication of code.

CLOVER [74] proposed by Liu *et al.* (2017), is a soft error detector incorporating two methods (a) sensor-based detectors [38] and (b) tail-DMR. In tail-DMR approach, compiler divides instruction sequence into two blocks (using control flow graphs) named head and tail. The head region depends on sensors to detect faults and the tail region performs selective duplication of instruction (DMR). This approach can detect and recover from soft errors only. CLOVER reports 46% reduction in instruction reduction when compared to full DMR. Program binary size increase in CLOVER is 36% whereas binary size increase for full-DMR is 86%. The Clover approach has a runtime overhead of 27% which is less compared to state-of-art fault detection techniques. The table (Table 5) shows the comparison of the above-mentioned techniques.

### 2.2.2 Built in Self-Test Methods

Software based Self-Test (SBST) [75] proposed by Psarakis *et al*. (2010), is a non-intrusive approach, which uses the instructions of the host machine to generate the test patterns thereby eliminating the need for extra test-related hardware support. Deployment of SBST on processors like MIPS R3000 [76], ARM4 core [77] and OpenRISC [78] achieves fault coverage of 95%, 94% and 93% respectively for permanent faults. SBST applied to bus based CMPs [79] conceptualized by Apostolakis *et al*. (2009), is a technique that deploys uniprocessor test programs on all cores for parallel execution, thereby reducing the total test execution time. It has a fault coverage of 91% for permanent faults. Multi-Threaded SBST (MT-SBST) [80] proposed by` Foutris *et al*. (2010), is an SBST approach extended with multithreading. The quickness in self-testing rate at core level (3.6x) and processor (6.0x) is observed in MT-SBST. 91% and 88% permanent fault coverage is observed for functional unit and for entire chip respectively in MT-SBST. The Access Control Extension (ACE) [81] proposed by Constantinides *et al*. (2007), has a special set of instructions that can access the state and control the microprocessor by periodically suspending execution. Special firmware is used to periodically suspend the microprocessor and execute ACE instructions to detect defects in the hardware. This approach achieves a fault coverage of 100% for permanent faults when implemented on eight SPARC cores of OpenSPARC T1 architecture.

### 2.2.3 Symptoms based Fault Detection Methods

The basic idea here is to detect suspicious activities (symptoms). Symptoms based fault detection techniques report faults in the software that is under observation for unpredictable activity. The symptoms that could result in faults are listed below:

(a) Data value like overflow or underflow values, values not matching with past or profiled values, and bit constants or invariants,

(b) Inconsistency in microarchitecture functions like page faults, thrashing, cache misses, and exceptions and

(c) Software behaviour anomalies, like suspension of normal activity by the operating system, fatal hardware traps, and abnormal termination of application.

Perturbation-Based Fault Screening (PBFS) [82] proposed by Racuans *et al*. (2007) detects faults due to occurrence of symptoms like overflow or underflow values, values not matching with past or profiled values, bit constants or invariants, etc. Restore [83] proposed by Wang and Patel (2006), is a technique that detects faults due to occurrence of symptoms like thrashing, page faults, Translation Lookaside Buffer (TLB) cache misses, exceptions, etc. SoftWare Anomaly Treatment (SWAT) [84] proposed by Man-Lap Li *et al*. (2008a), is a technique that detects faults due to occurrence of symptoms like abnormal termination of application and suspension of normal activity by the operating system.

In Restore proposed by Wang & Patel (2006), previous checkpoint is restored when a symptom is detected and the program re-execution begins from that point. Even false triggering of program re-execution in absence of soft error does not have any impact on the correctness. Frequent false triggering impacts performance. 93% of the injected faults are detected and corrected by Restore. It is equivalent to 2x rise in mean time between failures (MTBF) of the processor. Trace Based Fault Diagnosis (TBFD) [85] proposed by Man-Lap Li *et al*. (2008b), leverages the fault-free core in the system (identified by SWAT) to rerun and compare the results from faulty and fault-free cores. Fine-grained repair is activated after the fault is diagnosed. It successfully diagnoses over 98% of the faults to the faulty component. M-SWAT [86] proposed by Hari *et al*. (2009), incorporates redundancy (TMR) under necessary conditions for fault diagnosis. Fault diagnosis coverage is above 95%; faults propagated from faulty to fault-free core (Gizopoulos *et al*. 2011) [8] are successfully diagnosed.

**Table 6 Symptoms based fault detection techniques**

| Concept and year | Performance overhead | Area overhead | Target faults | Detection coverage | Detection latency |
|---|---|---|---|---|---|
| Restore (2006) | Low | Low | Transient fault | <100% | Unbounded |
| PBFS (2007) | Low | Low | Transient fault | <100% | Unbounded |
| SWAT (2008) | 5-15% | Low | Transient and permanent fault | 95% of permanent faults | Unbounded |
| TBFD (2008) | 5-15% | Low | Transient and Permanent fault | <100% | Unbounded |
| m-SWAT (2009) | 5-55% | Low | Transient and Permanent fault | <100% | Unbounded |
| Fault Hound (2015) | 10% | Low | Transient fault | 75% | Bounded |

FaultHound [36] proposed by Nitin *et al*. (2015) is a symptom-based fault tolerance method, which emphasises on false positives and recovery reduction. FaultHound monitors the locality of value by using filters. Filters use bias state machines to reduce false positives. The bias state machine keeps track of bit values toggling between "unchanged" and "change" state. Two unchanged occurrences are required to reach unchanged state. One occurrence of change is sufficient to move from unchanged to change state. This bias state machine improves fault coverage and reduces false positives. Experimental results report that FaultHound performs better than PBFS. 30% fault coverage and 8% false-positive rates are achieved by PBFS. FaultHound has 75% fault coverage and 3% false-positive rates with 25% energy overhead and 10% performance overhead. The concepts are itemized and summarized in the table (Table 6).

## 2.3 Reconfiguration and Repair techniques

StageNet Fabric [87] proposed by Gupta *et al*. (2008), is a first reconfigurable approach for multi-core systems with reliability as the key design criteria. In order to tolerate the permanent faults, StageNet constructs StageNetSlice (SNS) or logical processor core by selecting pipeline stages from the available pool of stages. StageNet performs reconfiguration at module level, pipeline stage level and core level. The pipeline segments in different cores forms a reconfigurable network. Each pipeline segment of a core is shared by other cores thereby providing redundancy. Every pipeline segment behaves as a processing element for every instruction yet to graduate. In order to keep track of register contents, StageNet maintains a scoreboard which occupies 1.4% of the total area. Experimental results reveal better MTTF at module level and pipeline stage level reconfiguration. Core re-configurability is a simpler technique to implement, and gives low returns in terms of lifetime extension with increasing defects. For embedded benchmarks, even after seven years, the Instruction per Cycle (IPC) of StageNet is 4x the baseline.

Detouring [33] proposed by Meixner *et al*. (2008), is a technique that generates a list of faults supplied to software translation unit which produces a code sequence which does not include the defective portion of the circuit. Detouring is provided for functional units, register file, and instruction cache, to obtain a fault-free core without hardware and performance overheads. Detouring has fault coverage of 42.5% and above 95% for permanent faults at the core level and instruction cache level respectively.

VIPER [88] proposed by Pellegrini *et al*. (2012), is a reconfigurable approach for an architecture with a fully distributed control logic. Instructions do not follow the path fixed at the design time as in a conventional architecture. VIPER is a service-oriented architecture consisting of hardware clusters. These hardware clusters are loosely connected to form execution engines. Hardware clusters offer one or more services. Faulty clusters can be disabled. In VIPER, a program is split into bundles and are assigned to hardware clusters. Instruction bundles can successfully execute as long as functional hardware clusters can provide all services needed for its instructions, in aggregate. Hardware clusters are connected in order to improve the reliability of the VIPER system. Built-in distributed control logic selects the hardware clusters to take part in any specific virtual pipeline at runtime. Fully distributed control logic enables VIPER to provide better performance in the presence of hundreds of permanent faults.

## 2.4 Micro-architectural Core Salvaging Techniques

Micro-architectural core salvaging is a method that disables the defective portion of the pipeline and exploits the redundancy by scheduling the operations on spare resources. Rescue [89] proposed by Schuchman and Vijaykumar (2005), is a core salvaging method, which addresses issues like (a) fault isolation time on core, (b) precision fault isolation in hardware structures and (c) commonly usable testing approach. Rescue provides Intra Cycle logic Independence (ICI), a concept to map out the faulty components where there exist scan detectable faults in the logic. The basic idea of ICI is that multiple faults can be simultaneously detected at the same time using one scan test vector such that multiple faults map to different components. When hard error count increases, Rescue has fewer cores disabled compared to core sparing (CS) technique. In CS technique, faulty cores on the CMP are disabled. The reason behind fewer cores impacted in Rescue is due to mapping of faulty components in logic by ICI. Yield-Adjusted instruction Throughput (YAT) is a metric that evaluates performance and yield together. Average YAT for Rescue improves over CS by 12% and 22% at 32nm and 18nm, respectively.

The Core Cannibalization Architecture (CCA) [90] proposed by Romanescu & Sorin (2008) is a technique that incorporates TMR or DMR configurations for cores. Cannibalized cores function normally in the absence of permanent faults. In the presence of permanent faults, Cannibalized cores will act as spare cores, or pipeline stages of a core can be cannibalized. CCA improves the lifetime reliability of a chip compared to chips without CCA. Implementation results show that for a period of 12 years, CCA for 3-core chips and 4-core chips experiences 63% and 64% rise in aggregated performance. If cores are part of TMR and DMR, 7% rise in performance is observed.

Srinivasan *et al*. (2005), conceptualized Structural Duplication (SD) [91] and Graceful Performance Degradation (GPD) methods. Graceful performance degradation is a method that tries to improve reliability by exploiting redundancy in the microarchitecture. Structural Duplication is a method of adding redundant micro-architectural components designated as spares. Spares are turned on when the original component fails. Performance degradation increases reliability by 1.42 times with less than 5% loss in performance. Structural duplication improves reliability by 3.17 times the base processor value. The base processor used in these approaches for simulation is 8-way, out-of-order POWER4 processor.

### 2.5 Architectural Salvaging Core Techniques

Core disabling, core sparing and core salvaging are the three design alternatives available to improve the lifetime reliability of the multicore systems. In the event of a defect in the core, defective cores are disabled in core disabling method, while spare cores are enabled in core sparing method. In core sparing method, spare cores not being used provide no economic returns or performance. Market sale price for core disabling is comparatively less due to a smaller number of cores with reduced performance. Core salvaging is a better alternative to core sparing and core disabling which allows defective cores to continue functioning.

Michael *et al*. (2009), proposed core salvaging [92] approach which exploits cross-core redundancy that exists naturally. This approach considers only permanent faults. The working principle of this approach is to migrate threads to another core that can execute this instruction thereby ignoring the defective core's disability to execute the same instruction in its microarchitecture. Architectural core salvaging approach provides fault coverage for larger die area and performance compared to core disabling and core sparing methods.

### 2.6 Error Recovery and Repair Techniques

Error is a manifestation of fault in the system. Error recovery can be done in two ways, they are forward error recovery (FER) and backward error recovery (BER). The principle followed in FER is to maintain redundant state information that allows to rebuild error-free state. Fail-over systems, DMR systems, TMR systems and pair-and-spare systems are four classes of FER approaches traditionally used. Marathon Endurance server [61] uses fail-over principles to recover from hardware error. Stratus ftserver [61] uses DMR systems for recovery. Overhead in FER is high as compared to BER. BER techniques, intermittently execute checkpoints to save present state, so that fault detection leads to automatic transition to previous valid state. SafetyNet [40] proposed by Sorin *et al*. (2002), is a technique that creates checkpoints at regular intervals for the system state, so that recovery to previous valid checkpoint (fault free) is executed. In SafetyNet, checkpoint interval is 100,000 cycles and logging happen only once in this interval. Hence, logging overhead is reduced. Pipelined checkpoint validation allows SafetyNet to tolerate long fault detection latency methods.

Revive I/O [42] proposed by Nakano *et al*. (2006) is an augmentation of Revive [41] proposed by Prvulovic *et al*. (2002). Both handle the output-commit problem, create global checkpoints and store the data related to checkpoints in memory. The requirement of output-commit problem is to communicate checkpoint validated fault-free data to outside of the sphere of recovery region. Log based rollback is the recovery mechanism followed in both SafetyNet and Revive approaches. Revive tolerates multiple transient faults in the system except in main memory and memory controller. Revive assumes that memory modules are safeguarded using distributed parity, and error detection mechanism is built-in. Revive can tolerate longer duration of fault detection latency of up to 100 ms, because it stores checkpoints and logs in main memory. SafetyNet can tolerate fault detection latency of up to 1 ms [8]. Performance overheads are high in Revive than SafetyNet.

Rebound [44] proposed by Rishi *et al*. (2011), is a model suitable for both multi-core and many-core systems that support directory-based cache coherence protocol. The interaction set (or set of processors) creates checkpoints and rollback all together. Their recovery operations are independent from other interaction sets. The rollback operation in Rebound is similar to Revive. The Rebound exploits the underneath coherence protocol to track inter-thread dependencies and record them. Rebound has been evaluated for 64 processors but Revive only for 16 processors. Rebound scales well on recovery latency, with a performance overhead at 2.5% (from 16 to 64 processors) and nominal increase in energy consumption.

Euripus [43] is an accelerator proposed by Doudalis & Prvulovic (2012) which supports bidirectional debugging, provides undo and redo-log checkpoints, which are created at different intervals. Since redo and undo logs are available, faster transition to previous valid state and moving forward to latest valid state are possible. Euripus incurs performance overhead is about 5% and roll-back latency is about 30%. Euripus incorporates hierarchical error recovery methods which empower it to tolerate high error rates with 99% efficiency.

All the above approaches that use redundancy-based solutions (at core/ thread/instruction/micro-architecture level), symptom-based fault detection methods, and micro-architecture / architecture core salving techniques to provide fault tolerance can be considered as classical approaches. These approaches incur area and power overheads. A non-classical approach by Upasani *et al*. (2014), uses acoustic sensor-based error detection and recovery technique [38] for soft errors. Fault is detected using sensors attached to the core die. This non-classical approach incurs less than 1% area overhead and 0.06% performance overhead compared to all the classical approaches.

The main objective of the sensor-based approach is to achieve zero Silent Data Corruption errors (SDC) per core and zero Detected Unrecoverable Errors (DUE) zero per core. This method equips the processor core with 68K sensor detectors to detect particle strikes that cause error in the cores to achieve zero SDC per core and zero DUE per core. This method has an error detection latency of 30-300 cycles at 2GHz.

As the alpha particles strike, sensor-based detectors would detect error and trigger recovery process. Checkpoints are incremental and periodic. The checkpoints creation includes (a) creating copies of architectural state (b) dirty lines of the cache are updated in main memory and (c) the program counter is saved. The checkpoint length is 2 million cycles. The checkpoints are scrutinized for validity. The recovery algorithm is executed to restore the architectural state and invalidate all dirty lines of L1 and Last level cache (LLC). This approach has 100% fault detection and recovery coverage.

## 3. CRITICAL EXAMINATION OF EXISTING LITERATURE

The merits and demerits of the techniques described above are summarized as follows:

(a) The redundant execution of every instruction must be minimized, but still fault tolerance should be provided. All the approaches provide redundant execution for every single instruction. The sphere of replication improves reliability with additional area and consumes extra power. The basic purpose of multi-core is to improve performance per watt per core which would not hold well in replication models.

(b) Fault detection at microarchitecture level targets transient faults. These techniques are not suitable for detecting permanent faults. Software based redundant techniques provide low cost, low intrusive hardware techniques which provide higher fault coverage for transient faults as high as 98%. Software based approaches focus more on transient faults detection and correction. In contrast, BIST techniques have high fault coverage for permanent faults. BIST techniques are not suitable for transient fault detection. The detection latency (100K cycles) is very high for fault symptom detection techniques. There is no guarantee that all faults would be detected by fault symptom detection-based techniques.

(c) Reconfigurable techniques improve MTTF at fine grain level but not coarse grain level. Significant improvement in lifetime reliability is seen in CCA, where part of a core is used only when needed. In CCA, performance overhead increases as the number of permanent faults increases.

(d) In architectural core salvaging, hardware-based thread migration approach exploits cross-core redundancy, such that cores are single Instruction Set Architecture (ISA) compliant. Architectural core salvaging is not a suitable approach for heterogeneous multi-core systems or many core systems that support multiple ISA or different ISA on cores.

(e) Among the error recovery techniques, Euripus is a bidirectional debugger. Unlike Euripus, Revive and Revive I/O, Rebound scales well when the number of cores increases. Scalability is evident in the experimental results of Rebound. Fault recovery latency is about 400ms for 16 core and increases to 820 ms for 64 cores with checkpoint interval of 4-5ms. Hence, Rebound is the most scalable recovery design for scalable multi-core systems.

(f) AR-SMT, DIVA and Slipstream forward all the values from leader core to trailing core. Only loads and branch information are forwarded to the trailing core in SRT, CRT, SRTR and CRTR. In RECVF, only critical values are forwarded. Performance overheads are very high in SRT and CRT. Idea of SlicK incorporated in SRT would result in 10% reduction in performance overhead. To some extent, both RECVF and RESEA are energy efficient redundant execution methods. Energy efficient redundant execution approaches are appropriate solutions for low power multi-core systems. The

**Table 7 Forwarded values from leader to trailing cores**

| Approach/methods | Forwarded values | Type of redundancy (core/thread level) |
|---|---|---|
| AR-SMT, | All values | Thread level |
| DIVA and | | Core level |
| Slipstream | | Core level |
| SRT, CRT, SRTR and CRTR | Load and Branches | (ALL)Thread level |
| RESEA & RECVF | Critical values | Core level |

table (Table 7) summarizes the concepts discussed above.

## 4. CLASSIFICATION OF FAULT TOLERANT MULTI-CORE SYSTEMS

The multi-core architecture with error resilience can be classified based on [a] Instruction set architecture [Single/Multiple] supported by cores, [b] microarchitecture of cores, [c] cores per socket [d] core hardened with codes [e] framework of thread migration. Considering above factors, we classify multi-core systems as (1) Homogeneous multi-core architectures with FT (2) Heterogeneous multi-core architectures with FT.

(4.1) Homogeneous multi-core architectures with fault tolerating model: In homogeneous multi-core systems, cores have identical microarchitecture and implement single Instruction set with symmetric performance, energy efficiency and resource set within each core. A thread migration-based fault tolerating model is proposed here. If a functional unit in a core fails to execute an instruction and fault is detected; thread executing the instruction is migrated to any other core connected to the same socket. The thread migration results in additional latency affecting the performance and increases the execution time of the instruction. The thread migration across the socket results in additional memory access time [not uniform] for the instruction that failed to execute on the parent core. Handling all failed instruction by a fixed core which provides fault free execution would be another design alternative. The thread migration model mentioned above can be explored in modern homogeneous multi-core architectures like 2 sockets with 4-core per socket Intel Xeon E5345 (Clovertown) provides uniform memory access time for all the cores, 2 socket with 2 core per socket AMD Opteron 2214 (Santa Rosa), 2 socket with 4-core per socket AMD Opteron 2214 (Santa Rosa), 2 socket with 8 core per socket Sun Ultra-Sparc T5140 T2+ (Victoria Falls) provides non-uniform memory access times. Considering Intel, AMD architectures and thread migration model, Single ISA Homogeneous Multi-core architecture can be further classified into: (a) Intra-socket fault tolerance by thread migration (b) Inter-socket fault tolerance by thread migration (c) Inter-socket fault tolerance by fixed core. The architectures considered for quantitative approach are shown in Figure 1(a) and (b).

(4.2) Heterogeneous multi-core architectures with FT

We classify them as (a) Single ISA heterogeneous cores with FT- minimal instruction set core (b) Single ISA heterogeneous leading-trailing cores (c) Multiple ISA heterogeneous cores FT-RIC (d) Multiple ISA superscalar core with Simple In-order [SIO] cores architecture (e) Multiple ISA heterogeneous cores with FT –minimal instruction set core.

(a) Single ISA heterogeneous cores with FT- minimal instruction set core

The Minimal instruction set [MIN- IS] is a subset of ISA supported by the cores in system. The MIN-IS core do not support all forms of instruction formats or addressing modes, as supported by other cores. Due permanent fault, instruction may fail to execute. The fault handler migrates the thread to MIN-IS core. The MIN-IS core emulates the instruction.

(b) Single ISA heterogeneous leading-trailing cores

Leading and trailing cores operate at different frequency and

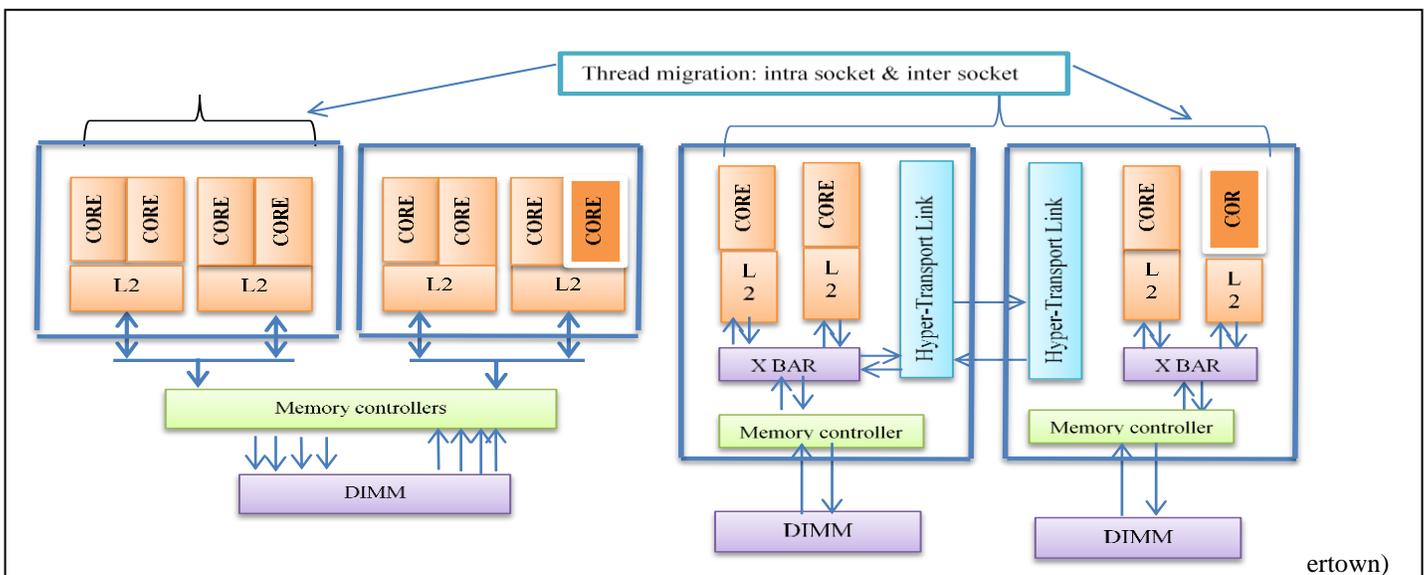

Figure 1 (a) Intel Xeon E5345 (Clovertown) (b) AMD Opteron 2214 (Santa Rosa)

voltage level, thereby exhibit performance asymmetry in different intervals of an application under experimentation. Both cores execute all the instructions of the workload at dissimilar operating frequencies thereby eliminating soft errors. Trailing cores captures all cache misses at L2 and have lesser memory penalty cycles, compared to leading cores. The lock stepped redundant execution on dynamically coupled cores [51] falls under this scheme. The leading and trailing threads loosely coupled, separated by slack [25], [62], [28] are used for error detection, thereby reduce checker overhead are conceptually same.

(c) Multiple ISA heterogeneous cores FT-RIC

The Reduced Instruction set Core [RIC] executes a one or few instructions equivalent to faulty instruction to obtain the desired result of the faulty instruction migrated to it. Our earlier work: One Instruction Core (OIC) [93], a variant of RIC, is a low power fault-tolerant small core enhances the reliability of multi-core systems. OIC executes only one instruction.

(d) Multiple ISA superscalar core with Simple In-order [SIO] cores architecture.

The parallel tasks of an application can be executed in SIOs and faulty instructions can be migrated to the big core. This core- eco system has a big core with higher performance, relatively low power SIOs and non-identical ISAs. Gschwind *et al*. [94] in CELL processor, core share same virtual memory and data types with difference in ISAs. The fault handler migrates the unsupported instructions to the respective cores. Finally, in terms of performance, power, memory access time and performance degradation stall cycles, heterogeneous core – ecosystem outperforms the homogeneous equivalents.

(e) Multiple ISA heterogeneous cores with FT –minimal instruction set

The Minimal Instruction Set [MIN- IS] is a subset of common instructions in different ISA supported by the cores in system. Operationally, MIN-IS described in (a) is applicable here. CPU-Graphics Processing Unit (GPU) systems can have a single MIN-IS core to provide fault tolerance.

## 5. ANALYTICAL APPROACH TO FAULT TOLERANT MODEL FOR HOMOGENEOUS MULTI-CORE SYSTEMS.

We discuss two fault tolerating model and they are (5.1) thread migration model and (5.2) process migration model. Fault tolerance aspects are quantitatively analyzed with respect to the architectures of Intel Xeon E5345 (Clovertown) and AMD Opteron 2214 (Santa Rosa) as shown in the figure 1(a) and 1 (b).

5.1 Thread migration model:

*Performance degradation stall cycles:* Performance degradation stall cycles is defined in this article as number of clock cycles spent in executing the Faulty Instruction (FI) on the receiving core, is inclusive of thread transfer time, CPU execution time$_{FI}$, memory access time and crossbar switch latency. Receiving core is assumed to be in active state.

Performance degradation stall cycles =

$\sum$ (Thread transfer time$_{2\text{-way}}$ + CPU execution time$_{FI}$ + Memory access time$_{FI}$ + Crossbar switch latency)
(1)

Thread transfer time$_{2\text{-way}}$ : number of clock cycles spent in resuming thread on the originated core. CPU execution time$_{FI}$: execution time in clock cycles for faulty instruction on new core. Memory access time$_{FI}$: memory access time for the faulty instruction on new core due miss in cache. Crossbar switch latency: clock delay in finding the route.

Equation (1) represents a light weight migration and is not applicable to process migration or heavy weight migration. Assuming two levels of cache, L1 private and L2 shared, operands for the faulty instruction would be available in L2 cache. Memory access time$_{FI}$ in Intra-socket fault tolerance by thread migration model (Figure 1 (a) with L2 cache shared and L2 cache exclusive is given by Equation (2) and (3) respectively.

Memory access time$_{FI}$ = Hit time$_{L1}$ + Miss rate$_{L1}$ x Hit time$_{L2}$
(2)

Memory access time$_{FI}$ = Hit time$_{L1}$ + Miss rate$_{L1}$ x (Hit time$_{L2}$ + Miss rate$_{L2}$ x (Miss penalty$_{L2}$)   (3)

Since operands are available in L2 cache, miss penalty for L2 cache is ignored in Equation 2. Miss penalty$_{L2}$ in the Equation (3) will be a uniform memory access time, since cores access memory which in not interleaved with respect to sockets present. Memory access time$_{FI}$ for Inter-socket fault tolerance by thread migration model in Figure 1 (b) with L2 cache exclusive, would include cross-bar latency also and is given by Equation (4).

Memory access time$_{FI}$ = Hit time$_{L1}$ + Miss rate$_{L1}$ x (Hit time$_{L2}$ + Miss rate$_{L2}$ x (Miss penalty$_{L2}$) +2 x cross-bar switch latency.
(4)

Here Miss penalty$_{L2}$ will be a non-uniform memory access time, since cores access memory which is interleaved on socket basis, which is unlikely in Intra-socket model. 2 x cross-bar switch latency will be the two latency for the thread that resumes operation on the originated core.

Arguably, the CPU execution time [95] can be rewritten as

CPU execution time = IC x (CPI + memory stall cycles + performance degradation stall cycles) x Clock cycle time
(5)

5.2 Process migration model:

When more than one functional unit in a core fails, process migration can be initiated from C1 to C2 core, assuming that C2 core is in active state. Process migrations time comprises two parts (a) overheads (clock cycles) and (b) state transfer and updating time. As soon as process migration begins, pipeline in C1 is flushed and C2 core is activated. Necessary updates like cache writeback and commit registers are performed in C1 and simultaneously C2 core is initialized for migration by activating necessary resources needed to accommodate new process from C1. If the C2 had been in completely shut down state or cold state [80], initialization would require additional one thousand cycles.

L2 to L2 cache transfer begins after the C2 is initialised. L2 to L2 cache transfer is not needed in case of intra socket fault tolerance by thread migration model where L2 cache is shared among cores. Branch predictor components (like branch target buffer, return address stack, conditional directional tables and target cache) contents transfer and register state transfer will consume significant portion of process migration time. Since new core branch predictor has to be trained with outcomes from other core (C1) for committed branch instructions to reduce the performance degradation. The normal execution of the process begins in C2 after the state transfer is completed and C1 core transits to permanent inactive state or further C1 would not be activated. Considering above discussion, we derive process migration time in Equation (6) as given below.

Process migration time = $\sum$ overheads + $\sum$ state transfer and updating time     (6)

$\sum$ Overheads includes (a) C2 initialization and cold state effects if applicable (b) pipeline flushing in C1 (c) training branch predictor components.

$\sum$ State transfer and updating time includes (a) L2 to L2 cache transfer (b) state transfer (C1 to C2) (c) cache writes backs and register updates (in C1).

5.3 Takeaways and Future directions

The analytical approach presented here will enhance the understanding of the readers about the architecture (Intel Xeon E5345 (Clover town), AMD Opteron 2214 (Santa Rosa)) with respect to performance degradation, memory access time and execution time. The sequence of events in thread migration and process migration will help coders (who develop simulators) to draw definite functional boundaries in estimating metrics with higher accuracy. Each individual functions in thread and process migrations helps coders to select appropriate data structures for implementation including runtime requirements with no ambiguity.

The reliability modelling and optimization problem can be formulated for the new design alternatives mentioned in section 4. Reliability analysis and modelling for the new design alternatives would facilitate development of new engineering applications. Thus, this article opens up many new directions for future in the area of low-cost fault tolerant reliable multi-core system.

## 6. CONCLUSION

This article has provided an understanding of error detection, error recovery, fault symptoms detection, software-based fault tolerant approaches, reconfigurable methods, micro-architectural and architectural salvaging approaches adapted for multi-core architectures. Gamut of fault tolerant methods presented in this article enhance reliability at different levels of abstraction for the architecture under consideration. Authors suggest new research directions by presenting new fault tolerant design alternatives for homogeneous and heterogeneous multi-core system that can further explored with simulation and analytical analysis for better understanding. Brief analytical approach for the suggested fault tolerating model would help both computer architects and developers enhance their understanding and explore the architecture with new analytical models.

## REFERNCES

[1] Moore, G.E., 1998. Cramming more components onto integrated circuits. *Proceedings of the IEEE*, *86*(1), pp.82-85.

[2] Moore, G.E., 2006. Lithography and the Future of Moore's Law. *IEEE Solid-State Circuits Society Newsletter*, *11*(3), pp.37-42.

[3] F Pollack. Pollack's rule of thumb for microprocessor and area. Available[online]:http://en.wikipedia.org/wiki/Pollack's_Rule.

[4] Dennard, R.H., Gaensslen, F.H., Yu, H.N., Rideout, V.L., Bassous, E. and LeBlanc, A.R., 1974. Design of ion-implanted MOSFET's with very small physical dimensions. *IEEE Journal of Solid-State Circuits*, *9*(5), pp.256-268.

[5] Tullsen, D.M., Eggers, S.J. and Levy, H.M., 1995, May. Simultaneous multithreading: Maximizing on-chip parallelism. In *Proceedings of the 22nd annual international symposium on Computer architecture* (pp. 392-403).

[6] Xbit Labs 2002, Intel Pentium 4 3.06 GHz CPU with hyper-threading technology: Killing two birds with astone…, Available[online]:http://www.xbitlabs.com/articles/cpu/display/pentium4-3066.html.

[7] Borkar, S., 2005. Designing reliable systems from unreliable components: the challenges of transistor variability and degradation. *IEEE Micro*, *25*(6), pp.10-16.

[8] Gizopoulos, D., Psarakis, M., Adve, S.V., Ramachandran, P., Hari, S.K.S., Sorin, D., Meixner, A., Biswas, A. and Vera, X., 2011, March. Architectures for online error detection and recovery in multicore processors. In *2011 Design, Automation & Test in Europe* (pp. 1-6). IEEE.

[9] Ray, J., Hoe, J.C. and Falsafi, B., 2001, December. Dual use of superscalar Datapath for transient-fault detection and recovery. In *Proceedings. 34th ACM/IEEE International Symposium on Microarchitecture. MICRO-34* (pp. 214-224). IEEE.

[10] Parashar, A., Gurumurthi, S. and Sivasubramaniam, A., 2004, June. A complexity-effective approach to Alu bandwidth enhancement for instruction-level temporal redundancy. In *Proceedings. 31st Annual International Symposium on Computer Architecture, 2004.* (pp. 376-386). IEEE.

[11] Nickel, J.B. and Somani, A.K., 2001, July. REESE: A method of soft error detection in microprocessors. In *2001 International Conference on Dependable Systems and Networks* (pp. 401-410). IEEE.

[12] Gomaa, M.A. and Vijaykumar, T.N., 2005, June. Opportunistic transient-fault detection. In *32nd International Symposium on Computer Architecture (ISCA'05)* (pp. 172-183). IEEE.

[13] Shyam, S., Constantinides, K., Phadke, S., Bertacco, V. and Austin, T., 2006. Ultra low-cost defect protection for microprocessor pipelines. *ACM SIGARCH Computer Architecture News*, *34*(5), pp.73-82.

[14] Meixner, A., Bauer, M.E. and Sorin, D., 2007, December. Argus: Low-cost, comprehensive error detection in simple cores. In *40th Annual IEEE/ACM International Symposium on Microarchitecture (MICRO 2007)* (pp. 210-222). IEEE.

[15] Hu, J.S., Link, G.M., John, J.K., Wang, S. and Ziavras, S.G., 2005, October. Resource-driven optimizations for transient-fault detecting superscalar microarchitectures. In *Asia-Pacific Conference on Advances in Computer Systems Architecture* (pp. 200-214). Springer, Berlin, Heidelberg.

[16] Soman, J., Miralaei, N., Mycroft, A. and Jones, T.M., 2015, October. REPAIR: Hard-error recovery via re-execution. In *2015 IEEE International Symposium on Defect and Fault Tolerance in VLSI and Nanotechnology Systems (DFTS)* (pp. 76-79). IEEE.

[17] Bernick, D., Bruckert, B., Vigna, P.D., Garcia, D., Jardine, R., Klecka, J. and Smullen, J., 2005, June. NonStop advanced architecture. In *2005 International Conference on Dependable Systems and Networks (DSN'05)* (pp. 12-21). IEEE.

[18] Austin, T.M., 1999, November. DIVA: A reliable substrate for deep submicron microarchitecture design. In *MICRO-32. Proceedings of the 32nd Annual ACM/IEEE International Symposium on Microarchitecture* (pp. 196-207). IEEE.

[19] Purser, Z., Sundaramoorthy, K. and Rotenberg, E., 2000, December. A study of slipstream processors. In *Proceedings of the 33rd annual ACM/IEEE International Symposium on Microarchitecture* (pp. 269-280).

[20] Rashid, M.W., Tan, E.J., Huang, M.C. and Albonesi, D.H., 2005, September. Exploiting coarse-grain verification parallelism for power-efficient fault tolerance. In *14th International Conference on Parallel Architectures and Compilation Techniques (PACT'05)* (pp. 315-325). IEEE.

[21] Li, H.T., Chou, C.Y., Hsieh, Y.T., Chu, W.C. and Wu, A.Y., 2017. Variation-aware reliable many-core system design by exploiting inherent core redundancy. *IEEE Transactions on Very Large Scale Integration (VLSI) Systems*, *25*(10), pp.2803-2816.

[22] Iturbe, X., Venu, B., Penton, J. and Ozer, E., 2017, October. A" high resilience" mode to minimize soft error vulnerabilities in ARM cortex-R CPU pipelines: work-in-progress. In *Proceedings of the 2017 International Conference on Compilers, Architectures and Synthesis for Embedded Systems Companion* (pp. 1-2).

[23] Ainsworth, S. and Jones, T.M., 2018, June. Parallel error detection using heterogeneous cores. In *2018 48th Annual IEEE/IFIP International Conference on Dependable Systems and Networks (DSN)* (pp. 338-349). IEEE.

[24] Spainhower, L. and Gregg, T.A., 1999. IBM S/390 parallel enterprise server G5 fault tolerance: A historical perspective. *IBM Journal of Research and Development*, *43*(5.6), pp.863-873.

[25] Rotenberg, E., 1999, June. AR-SMT: A microarchitectural approach to fault tolerance in microprocessors. In *Digest of Papers. Twenty-Ninth Annual International Symposium on Fault-Tolerant Computing (Cat. No. 99CB36352)* (pp. 84-91). IEEE.

[26] Reinhardt, S.K. and Mukherjee, S.S., 2000, June. Transient fault detection via simultaneous multithreading. In *Proceedings of 27th International Symposium on Computer Architecture (IEEE Cat. No. RS00201)* (pp. 25-36). IEEE.

[27] Vijaykumar, T.N., Pomeranz, I. and Cheng, K., 2002, May. Transient-fault recovery using simultaneous multithreading. In *Proceedings 29th Annual International Symposium on Computer Architecture* (pp. 87-98). IEEE.

[28] Gomaa, M., Scarbrough, C., Vijaykumar, T.N. and Pomeranz, I., 2003, June. Transient-fault recovery for chip multiprocessors. In *30th Annual International Symposium on Computer Architecture, 2003. Proceedings.* (pp. 98-109). IEEE.

[29] Huang, B., Sass, R., Debardeleben, N. and Blanchard, S., 2014, June. Harnessing unreliable cores in heterogeneous architecture: The PyDac programming model and runtime.

In *2014 44th Annual IEEE/IFIP International Conference on Dependable Systems and Networks* (pp. 744-749). IEEE.

[30] Oh, N., Shirvani, P.P. and McCluskey, E.J., 2002a. Error detection by duplicated instructions in super-scalar processors. *IEEE Transactions on Reliability*, 51(1), pp.63-75.

[31] Kanawati, G.A., Nair, V.S., Krishnamurthy, N. and Abraham, J.A., 1996, September. Evaluation of integrated system-level checks for on-line error detection. In *Proceedings of IEEE International Computer Performance and Dependability Symposium* (pp. 292-301). IEEE.

[32] Reis, G.A., Chang, J., Vachharajani, N., Rangan, R. and August, D.I., 2005, March. SWIFT: Software implemented fault tolerance. In *International symposium on Code generation and optimization* (pp. 243-254). IEEE.

[33] Meixner, A. and Sorin, D.J., 2008, June. Detouring: Translating software to circumvent hard faults in simple cores. In *2008 IEEE International Conference on Dependable Systems and Networks with FTCS and DCC (DSN)* (pp. 80-89). IEEE.

[34] Li, M.L., Ramachandran, P., Sahoo, S.K., Adve, S.V., Adve, V.S. and Zhou, Y., 2008b, June. Trace-based microarchitecture-level diagnosis of permanent hardware faults. In *2008 IEEE International Conference on Dependable Systems and Networks with FTCS and DCC (DSN)* (pp. 22-31). IEEE.

[35] Hari, S.K.S., Li, M.L., Ramachandran, P., Choi, B. and Adve, S.V., 2009, December. mSWAT: Low-cost hardware fault detection and diagnosis for multicore systems. In *2009 42nd Annual IEEE/ACM International Symposium on Microarchitecture (MICRO)* (pp. 122-132). IEEE.

[36] Nitin, Irith Pomeranz and T.N. Vijaykumar, 2015, FaultHound: Value-Locality-Based Soft-Fault Tolerance. In Proceedings of the forty-second Annual ACM/IEEE International Symposium on Computer Architecture, pp. 668-681.

[37] Liu, Q., Jung, C., Lee, D. and Tiwari, D., 2016. Compiler-directed soft error detection and recovery to avoid DUE and SDC via Tail-DMR. *ACM Transactions on Embedded Computing Systems (TECS)*, 16(2), pp.1-26.

[38] Upasani, G., Vera, X. and González, A., 2014. Avoiding core's due & sdc via acoustic wave detectors and tailored error containment and recovery. *ACM SIGARCH Computer Architecture News*, 42(3), pp.37-48.

[39] Mahmoud, A., Venkatagiri, R., Ahmed, K., Misailovic, S., Marinov, D., Fletcher, C.W. and Adve, S.V., 2019, April. Minotaur: Adapting software testing techniques for hardware errors. In *Proceedings of the Twenty-Fourth International Conference on Architectural Support for Programming Languages and Operating Systems* (pp. 1087-1103).

[40] Sorin, D.J., Martin, M.M., Hill, M.D. and Wood, D.A., 2002, May. SafetyNet: Improving the availability of shared memory multiprocessors with global checkpoint/recovery. In *Proceedings 29th Annual International Symposium on Computer Architecture* (pp. 123-134). IEEE.

[41] Prvulovic, M., Zhang, Z. and Torrellas, J., 2002. ReVive: Cost-effective architectural support for rollback recovery in shared-memory multiprocessors. *ACM SIGARCH Computer Architecture News*, 30(2), pp.111-122.

[42] Nakano, J., Montesinos, P., Gharachorloo, K. and Torrellas, J., 2006, February. ReViveI/O: Efficient handling of I/O in highly-available rollback-recovery servers. In *The Twelfth International Symposium on High-Performance Computer Architecture, 2006.* (pp. 200-211). IEEE.

[43] Doudalis, I. and Prvulovic, M., 2012, June. Euripus: A flexible unified hardware memory checkpointing accelerator for bidirectional-debugging and reliability. In *2012 39th Annual International Symposium on Computer Architecture (ISCA)* (pp. 261-272). IEEE.

[44] Agarwal, R., Garg, P. and Torrellas, J., 2011, June. Rebound: scalable checkpointing for coherent shared memory. In *Proceedings of the 38th annual international symposium on Computer architecture* (pp. 153-164).

[45] Sarangi, S.R., Greskamp, B. and Torrellas, J., 2006, June. Cadre: Cycle-accurate deterministic replay for hardware debugging. In *International Conference on Dependable Systems and Networks (DSN'06)* (pp. 301-312). IEEE.

[46] X W. Bartlett and B. Ball,1998. Tandems approach to fault tolerance. Tandem Systems, 4(1), pp.84-95.

[47] Fair, M.L., Conklin, C.R., Swaney, S.B., Meaney, P.J., Clarke, W.J., Alves, L.C., Modi, I.N., Freier, F., Fischer, W. and Weber, N.E., 2004. Reliability, Availability, and Serviceability (RAS) of the IBM eServer z990. *IBM Journal of Research and Development*, 48(3.4), pp.519-534.

[48] Aggarwal, N., Ranganathan, P., Jouppi, N.P. and Smith, J.E., 2007. Configurable isolation: building high availability systems with commodity multi-core processors. *ACM SIGARCH Computer Architecture News*, 35(2), pp.470-481.

[49] Smolens, J.C., Gold, B.T., Kim, J., Falsafi, B., Hoe, J.C. and Nowatzyk, A.G., 2004. Fingerprinting: Bounding soft-error detection latency and bandwidth. *ACM SIGOPS Operating Systems Review*, 38(5), pp.224-234.

[50] Smolens, J.C., Gold, B.T., Falsafi, B. and Hoe, J.C., 2006, December. Reunion: Complexity-effective multicore redundancy. In *2006 39th Annual IEEE/ACM International Symposium on Microarchitecture (MICRO'06)* (pp. 223-234). IEEE.

[51] LaFrieda, C., Ipek, E., Martinez, J.F. and Manohar, R., 2007, June. Utilizing dynamically coupled cores to form a

resilient chip multiprocessor. In *37th Annual IEEE/IFIP International Conference on Dependable Systems and Networks (DSN'07)* (pp. 317-326). IEEE.

[52] Sundaramoorthy, K., Purser, Z. and Rotenberg, E., 2000. Slipstream processors: Improving both performance and fault tolerance. *ACM SIGPLAN Notices*, 35(11), pp.257-268.

[53] Subramanyan, P, Singh, V, Saluja, KK and Larsson, E ,2009. Power-Efficient Redundant Execution for Chip Multiprocessors. In Proceedings *of IEEE 3rd workshop on Dependable and Secure Nano computing held in conjunction with IEEE DSN 2009*, pp. 1-6.

[54] Subramanyan, P, Singh, V, Saluja, KK and Larsson, E,2010. Energy-Efficient Redundant Execution for Chip Multiprocessors. In *Proceedings of the twentieth ACM Great Lakes Symposium on VLSI*, pp.143-146.

[55] Subramanyan, P, Singh V, KK, Saluja & Larsson, E,2010. Energy-Efficient Fault Tolerance in Chip Multiprocessors Using Critical Value Forwarding. In *Proceedings of IEEE International conference on Dependable Systems and Networks,* pp.23-35.

[56] Gopalakrishnan, S. and Singh, V., 2017, October. REMORA: a hybrid low-cost soft-error reliable fault tolerant architecture. In *2017 IEEE International Symposium on Defect and Fault Tolerance in VLSI and Nanotechnology Systems (DFT)* (pp. 1-6). IEEE.

[57] Soman, J. and Jones, T.M., 2017, October. High performance fault tolerance through predictive instruction re-execution. In *2017 IEEE International Symposium on Defect and Fault Tolerance in VLSI and Nanotechnology Systems (DFT)* (pp. 1-4). IEEE.

[58] Ainsworth, S. and Jones, T.M., 2018, June. Parallel error detection using heterogeneous cores. In *2018 48th Annual IEEE/IFIP International Conference on Dependable Systems and Networks (DSN)* (pp. 338-349). IEEE.

[59] Smolens, J.C., Kim, J., Hoe, J.C. and Falsafi, B., 2004, December. Efficient resource sharing in concurrent error detecting superscalar microarchitectures. In *37th International Symposium on Microarchitecture (MICRO-37'04)* (pp. 257-268). IEEE.

[60] Vera, X., Abella, J., Carretero, J. and González, A., 2010. Selective replication: A lightweight technique for soft errors. *ACM Transactions on Computer Systems (TOCS)*, 27(4), pp.1-30.

[61] Mukherjee, S, Architecture design for soft errors, Morgan Kaufmann, 2011.

[62] Mukherjee, S.S., Kontz, M. and Reinhardt, S.K., 2002, May. Detailed design and evaluation of redundant multi-threading alternatives. In *Proceedings 29th annual international symposium on computer architecture* (pp. 99-110). IEEE.

[63] Parashar, A., Sivasubramaniam, A. and Gurumurthi, S., 2006. SlicK: slice-based locality exploitation for efficient redundant multithreading. *ACM SIGOPS Operating Systems Review*, 40(5), pp.95-105.

[64] Kumar, S. and Aggarwal, A., 2008, February. Speculative instruction validation for performance-reliability trade-off. In *2008 IEEE 14th International Symposium on High Performance Computer Architecture* (pp. 405-414). IEEE.

[65] Huang, B., Sass, R., Debardeleben, N. and Blanchard, S., 2014, June. Harnessing unreliable cores in heterogeneous architecture: The PyDac programming model and runtime. In *2014 44th Annual IEEE/IFIP International Conference on Dependable Systems and Networks* (pp. 744-749). IEEE.

[66] Schuette, M.A. and Shen, J.P., 1987. Processor control flow monitoring using signatured instruction streams. *IEEE Transactions on Computers*, 36(3), pp.264-276.

[67] Namjoo, M., Techniques for Concurrent Testing of VLSI Processor. In *Proc. of the International Test Conference (ITC)* (pp. 416-468).

[68] Wilken, K. and Shen, J.P., 1990. Continuous signature monitoring: low-cost concurrent detection of processor control errors. *IEEE Transactions on Computer-Aided Design of Integrated Circuits and Systems*, 9(6), pp.629-641.

[69] Oh, N., Shirvani, P.P. and McCluskey, E.J., 2002b. Control-flow checking by software signatures. *IEEE transactions on Reliability*, 51(1), pp.111-122.

[70] Oh, N., Shirvani, P.P. and McCluskey, E.J., 2002a. Error detection by duplicated instructions in super-scalar processors. *IEEE Transactions on Reliability*, 51(1), pp.63-75.

[71] Reis, G.A., Chang, J., Vachharajani, N., Mukherjee, S.S., Rangan, R. and August, D.I., 2005, June. Design and evaluation of hybrid fault-detection systems. In *32nd International Symposium on Computer Architecture (ISCA'05)* (pp. 148-159). IEEE.

[72] Wang, C., Kim, H.S., Wu, Y. and Ying, V., 2007, March. Compiler-managed software-based redundant multi-threading for transient fault detection. In *International Symposium on Code Generation and Optimization (CGO'07)* (pp. 244-258). IEEE.

[73] Chang, J., Reis, G.A. and August, D.I., 2006, June. Automatic instruction-level software-only recovery. In *International Conference on Dependable Systems and Networks (DSN'06)* (pp. 83-92). IEEE.

[74] Liu, Q., Jung, C., Lee, D. and Tiwari, D., 2016. Compiler-directed soft error detection and recovery to avoid

DUE and SDC via Tail-DMR. *ACM Transactions on Embedded Computing Systems (TECS)*, *16*(2), pp.1-26.

[75] Psarakis, M., Gizopoulos, D., Sanchez, E. and Reorda, M.S., 2010. Microprocessor software-based self-testing. *IEEE Design & Test of Computers*, *27*(3), pp.4-19.

[76] Kranitis, N., Paschalis, A., Gizopoulos, D. and Xenoulis, G., 2007. Software-based self-testing of embedded processors. In *Processor design* (pp. 447-481). Springer, Dordrecht.

[77] Chen, C.H., Wei, C.K., Lu, T.H. and Gao, H.W., 2007. Software-based self-testing with multiple-level abstractions for soft processor cores. *IEEE transactions on very large scale integration (VLSI) systems*, *15*(5), pp.505-517.

[78] Gizopoulos, D., Psarakis, M., Hatzimihail, M., Maniatakos, M., Paschalis, A., Raghunathan, A. and Ravi, S., 2008. Systematic software-based self-test for pipelined processors. IEEE Transactions on Very Large Scale Integration (VLSI) Systems, 16(11), pp.1441-1453.

[79] Apostolakis, A., Gizopoulos, D., Psarakis, M. and Paschalis, A., 2009. Software-based self-testing of symmetric shared-memory multiprocessors. *IEEE Transactions on Computers*, *58*(12), pp.1682-1694.

[80] Foutris, N., Psarakis, M., Gizopoulos, D., Apostolakis, A., Vera, X. and González, A., 2010, November. MT-SBST: Self-test optimization in multithreaded multicore architectures. In *2010 IEEE International Test Conference* (pp. 1-10). IEEE.

[81] Constantinides, K., Mutlu, O., Austin, T. and Bertacco, V., 2007, December. Software-based online detection of hardware defects mechanisms, architectural support, and evaluation. In *40th Annual IEEE/ACM International Symposium on Microarchitecture (MICRO 2007)* (pp. 97-108). IEEE.

[82] Racunas, P., Constantinides, K., Manne, S. and Mukherjee, S.S., 2007, February. Perturbation-based fault screening. In *2007 IEEE 13th International Symposium on High Performance Computer Architecture* (pp. 169-180). IEEE.

[83] Wang, N.J. and Patel, S.J., 2006. ReStore: Symptom-based soft error detection in microprocessors. *IEEE Transactions on Dependable and Secure Computing*, *3*(3), pp.188-201.

[84] Li, M.L., Ramachandran, P., Sahoo, S.K., Adve, S.V., Adve, V.S. and Zhou, Y., 2008a. Understanding the propagation of hard errors to software and implications for resilient system design. *ACM SIGPLAN Notices*, *43*(3), pp.265-276.

[85] Li, M.L., Ramachandran, P., Sahoo, S.K., Adve, S.V., Adve, V.S. and Zhou, Y., 2008b, June. Trace-based microarchitecture-level diagnosis of permanent hardware faults. In *2008 IEEE International Conference on Dependable Systems and Networks with FTCS and DCC (DSN)* (pp. 22-31). IEEE.

[86] Hari, S.K.S., Li, M.L., Ramachandran, P., Choi, B. and Adve, S.V., 2009, December. mSWAT: Low-cost hardware fault detection and diagnosis for multicore systems. In *2009 42nd Annual IEEE/ACM International Symposium on Microarchitecture (MICRO)* (pp. 122-132). IEEE.

[87] Gupta, S., Feng, S., Ansari, A., Blome, J. and Mahlke, S., 2008, November. The StageNet fabric for constructing resilient multicore systems. In *2008 41st IEEE/ACM International Symposium on Microarchitecture* (pp. 141-151). IEEE.

[88] Pellegrini, A., Greathouse, J.L. and Bertacco, V., 2012. Viper: virtual pipelines for enhanced reliability. *ACM SIGARCH Computer Architecture News*, *40*(3), pp.344-355.

[89] Schuchman, E. and Vijaykumar, T.N., 2005, June. Rescue: A microarchitecture for testability and defect tolerance. In *32nd International Symposium on Computer Architecture (ISCA'05)* (pp. 160-171). IEEE.

[90] Romanescu, B.F. and Sorin, D.J., 2008, October. Core cannibalization architecture: improving lifetime chip performance for multicore processors in the presence of hard faults. In *2008 International Conference on Parallel Architectures and Compilation Techniques (PACT)* (pp. 43-51). IEEE.

[91] Srinivasan, J., Adve, S.V., Bose, P. and Rivers, J.A., 2005, June. Exploiting structural duplication for lifetime reliability enhancement. In *32nd International Symposium on Computer Architecture (ISCA'05)* (pp. 520-531). IEEE.

[92] Powell, M.D., Biswas, A., Gupta, S. and Mukherjee, S.S., 2009. Architectural core salvaging in a multi-core processor for hard-error tolerance. *ACM SIGARCH Computer Architecture News*, *37*(3), pp.93-104.

[93] Shashikiran. Venkatesha and R. Parthasarathi, 2019. 32-Bit One Instruction Core: A Low-Cost, Reliable, and Fault-Tolerant Core for Multicore Systems. Journal of Testing and Evaluation 47, no. 6, pp. 3941–3962.

[94] Gschwind, M., Hofstee, H.P., Flachs, B., Hopkins, M., Watanabe, Y. and Yamazaki, T., 2006. Synergistic processing in cell's multicore architecture. *IEEE micro*, *26*(2), pp.10-24.

[95] Hennessy, J.L. and Patterson, D.A., 2011. *Computer architecture: a quantitative approach*. Elsevier.

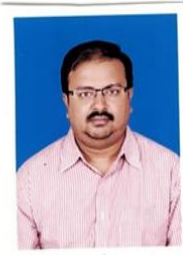

**Shashikiran Venkatesha** has completed his PhD from Department of Information science and Technology, College of Engineering Guindy, Anna University Chennai, India. He has completed his Bachelor's degree and Master's in Computer science and Engineering with distinction from University of Madras and Anna University respectively. His research interest is Digital Systems Engineering, Digital arithmetic, Digital Testing System and Testability designs, Microprocessor and Microcontroller's interfacing, Computer architecture and Reliability. Presently, he is working as a Assistant Professor – Senior Grade at Vellore Institute of Technology.

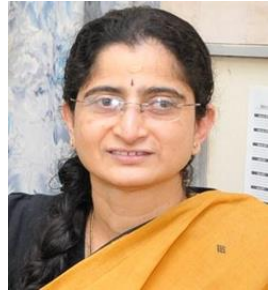

**Ranjani Parthasarathi** is Professor at Department of Information science and Technology, College of engineering Guindy Anna University Chennai India. She has completed Bachelor's degree in Electronics and Communication Engineering from Anna University and Master's from Illinois Institute of Technology Chicago. She did her PhD from Indian Institute of Technology Madras India. She has 30 years of teaching experience and published papers in areas related Computer architecture, Natural Language processing, Parallel architecture and Artificial Intelligence.